\documentstyle{article}

\newcommand{\ba}{\begin{eqnarray}}
\newcommand{\ea}{\end{eqnarray}}
\newcommand{\ii}{\'\i}
\begin{document}
\pagestyle{empty}

\vbox{\vspace{1cm}}

\noindent
\begin{center}{\bf\uppercase{A Symmetry Adapted Approach\\
to Vibrational Excitations \\ 
in Atomic Clusters}}
\end{center}

\vspace{10mm}
\noindent\hskip 1in \vbox{\noindent
A. Frank$^{1,2}$, R. Lemus$^{1}$, R. Bijker$^{1}$, 
F. P\'erez-Bernal$^{3}$, \\ 
and J. M. Arias$^{3}$\vskip5mm \noindent
$^{1}$Instituto de Ciencias Nucleares, U.N.A.M.,\\
$\hphantom{^1}$A.P. 70-543, 04510 M\'exico D.F., M\'exico\\
$^{2}$Instituto de F\'{\i}sica, Laboratorio de Cuernavaca,\\
$\hphantom{^1}$A.P. 139-B, Cuernavaca, Morelos, M\'exico\\
$^{3}$Departamento de F\'{\i}sica At\'omica, Molecular y Nuclear,\\
$\hphantom{^1}$Facultad de F\'{\i}sica, Universidad de Sevilla,\\
$\hphantom{^1}$Apdo. 1065, 41080 Sevilla, Espa\~na
}

\vspace{10mm}

\begin{abstract}
An algebraic method especially suited to describe strongly anharmonic
vibrational spectra in molecules may be an appropriate framework to
study vibrational spectra of Na$^+_n$ clusters, where nearly flat
potential energy surfaces and the appearance of close lying isomers
have been reported.  As an illustration we describe the model and 
apply it to the Be$_4$, H$_3^+$, Be$_3$ and Na$_3^+$ clusters.
\end{abstract}

\vspace{8mm}
\noindent{\bf INTRODUCTION}
\vspace{6mm}

The study of metallic clusters has been mainly concerned with the
electronic properties of these systems.  It has been suggested,
however, that in analogy to the way crystalline structure determines
the optical response and metallic properties in solid-state physics,
the geometric structure may play a  significant role in the case of
cluster physics \cite{uno,dos}. Small finite size systems present special
problems which cannot directly be dealt with by means of the methods
 applied to the bulk.  Nuclear theorists, on the other
hand, have much experience dealing with ``the many body problem'' and the
application  of these methods to clusters is already leading to
significant contributions to their understanding.  These techniques,
however, usually ignore the position of atomic nuclei \cite{tres}.  In
\cite{uno,dos} ab initio calculations for Na$^+_n (n = 2-9,11,21)$
clusters at low temperature are reported, which allow the assignment
of specific cluster geometries by comparing the theoretical results
with experimental optical depletion spectra for these
systems.   While for dimers and trimers the rovibrational
spectra lead to a precise determination of internuclear distances and
ground-state potential energy surfaces (P.E.S.), up to now the
resolution of the vibronic structure for $n\geq 4$ is not enough to
carry out a theoretical analysis that 
  would confirm the results of ref. [2] and
give information on the P.E.S. for the heavier clusters \cite{cuatro}.  Even
if the rovibrational data become available in the near future, the
calculations of [2] imply that the corresponding potential surfaces
are in general quite flat and thus one expects anharmonic effects to
be significant.  Also, in a number of cases close lying isomeric
structures are found \cite{dos}.  Since the usual molecular physics
calculations are well suited for deep minima and not for ``soft''
molecules, it is important to consider alternative models in order to
analyse the experimental results.  In this paper we describe a new
model which was designed to incorporate anharmonic behavior from the
outset and which has been applied to a number of molecules, including
the Na$^+_3$ cluster \cite{cinco}-\cite{ocho}.  We illustrate the method with
applications to  some $D_{3h}$ molecules and clusters and to the
Be$_4$ cluster.

\vspace{8mm}
\noindent{\bf THE U(2) VIBRON MODEL}
\vspace{6mm}

The   model  is based on  the isomorphism of the $U(2)$ Lie
algebra and the one dimensional Morse oscillator,
 whose eigenstates  can be associated with  $U(2)\supset
 SO(2)$ states \cite{nueve}.   
In the framework of the model the total 
number of bosons $N$ is fixed by the potential shape and the
eigenvalue $m$   of the $SO(2)$ generator $J_z$,
takes the values $m = \pm N/2$, $\pm (N-2)/2, \dots$.  The Morse
spectrum is reproduced twice and consequently for these applications
the $m$-values must be restricted to be positive.  In terms of the
  $U(2)$ algebra,  the Morse
Hamiltonian has the algebraic realization 
\ba
\hat  H = A ({\bf J}^2 - J^2_z) ~~.
\ea
with eigenvalues 
\ba
E_M =AN [ (v+1/2) - v^2/N] ~~ , 
\ea
where the label $v=j-m$ denotes the number of quanta in the
oscillator. 
  The parameters $N$ and $A$ appearing in (2) are
related with the usual  
 harmonic and anharmonic constants $\omega_e$ and
$x_e\omega_e$ used in spectroscopy \cite{diez}. 
 We now  consider the \ $U_i(2) \ \supset \ SU_i(2) \ \supset \ SO_i(2)$ 
\ algebra, \ generated \ by \ the \ set \  
$\{ \hat G_i \} \equiv $  $ \{ \hat N_i, \, \hat J_{+,i}, 
\, \hat J_{-,i}, \, \hat J_{0,i} \}$, satisfying the commutation 
relations 
\ba
\, [ \hat J_{0,i}, \hat J_{\pm,i}] \;=\; \pm \hat J_{\pm,i} ~,
\hspace{1cm} 
\, [ \hat J_{+,i}, \hat J_{-,i}] \;=\; 2 \hat J_{0,i} ~,
\hspace{1cm} 
\, [ \hat N_i, \hat J_{\mu,i}] \;=\; 0 ~, \label{jmui}
\ea
with $\mu=\pm,0$.  
 For the symmetric irreducible representation
$[N_i,0]$  
of $U_i(2)$ one can show that the Casimir operator is given by
 \cite{ocho} 
$\vec{J}_i^{\, 2} = \hat N_i(\hat N_i+2)/4$, 
from which follows the identification $j_i=N_i/2$. 

In the algebraic approach each relevant interatomic interaction
is associated with a $U_i(2)$ algebra \cite{cinco,seis}. As a first 
example, we  
consider the Be$_4$ cluster, which has a tetrahedral shape.  ${\cal
D}_{3h}$ molecules can be similarly treated.  
 In the Be$_4$  
case there are six $U_i(2)$ algebras involved ($i=1,\ldots,6$).
 The operators in the model are expressed in terms of the 
generators of these algebras, and the symmetry requirements of the 
tetrahedral group ${\cal T}_d$  can be readily imposed
 \cite{seis,once}. 
The local operators $\{ \hat G_i \}$
acting on bond $i$ can be projected to any of the fundamental 
irreps $\Gamma=A_1$, $E$ and  $F_2$.
Using the $\hat J_{\mu,i}$ generators (3)
we obtain the ${\cal T}_d$ tensors
\ba
\hat T^{\Gamma}_{\mu,\gamma} &=& 
\sum_{i=1}^{6} \, \alpha^{\Gamma}_{\gamma,i} \, \hat J_{\mu,i} ~,
\ea
where $\mu=\pm,0$ and $\gamma$ denotes the component of $\Gamma$. 
The expansion coefficients are the same as those given in the one
phonon wave functions \cite{siete}. 
 The Hamiltonian operator can be constructed by repeated couplings 
of these tensors to a total symmetry $A_1$, since it must commute 
with all operations in ${\cal T}_d$. This is accomplished by means
of the ${\cal T}_d$-Clebsch-Gordan coefficients
 \cite{seis,once,doce}. 

All calculations can be carried out in a symmetry-adapted basis, which 
is projected from the local basis
\ba
\begin{array}{ccccccccccccc}
U_1(2) &\otimes& \cdots &\otimes& U_6(2) &\supset& 
SO_1(2) &\otimes& \cdots &\otimes& SO_6(2) &\supset& SO(2) \\
\downarrow && && \downarrow && \downarrow && && 
\downarrow && \downarrow \\ 
| \;\; [N_1] &,& \ldots &,& [N_6] &;& 
v_1 &,& \ldots &,& v_6 &;& \; V \;\; \rangle 
\end{array}
\ea
in which each anharmonic oscillator is well defined. By 
symmetry  
considerations, $N_i=N$ for the six oscillators, $v_i$, 
denotes  
the  number of quanta in bond $i$ and $V=\sum_i v_i$ is the total 
number of quanta \cite{trece,ocho}.    
The local basis states for each 
oscillator are usually written as $|N_i,v_i \rangle$, where 
$v_i=(N_i-2m_i)/2=0,1, \ldots [N_i/2]$ denotes the number of oscillator 
quanta in the $i$-th oscillator.  
The  phonon states $|^{V}\phi^{\Gamma}_{\gamma}>$ can  be 
  constructed using the Clebsch-Gordan coefficients of 
${\cal T}_d$ \cite{seis,once}. Since all operators are expressed in
terms of  
powers of the $U_i(2)$ generators, their matrix elements can 
be easily  evaluated in closed form. The symmetry-adapted operators 
(11) and states \cite{siete} are the building blocks of the
model.

We now proceed to explicitly construct the  Be$_4$   Hamiltonian. 
For interactions that are at most quadratic in the 
generators the  procedure yields 
\ba
\hat H_0 &=& \omega_1 \, \hat{\cal H}_{A_1} 
+ \omega_2 \, \hat{\cal H}_{E} 
+ \omega_3 \, \hat{\cal H}_{F_2} 
+ b_2 \, \hat{\cal V}_{E} + b_3 \, \hat{\cal V}_{F_2} ~, \label{H0}
\ea
with
\ba
\hat{\cal H}_{\Gamma} &=& \frac{1}{2N} \sum_{\gamma} \left( 
  \hat T^{\Gamma}_{-,\gamma} \, \hat T^{\Gamma}_{+,\gamma}
+ \hat T^{\Gamma}_{+,\gamma} \, \hat T^{\Gamma}_{-,\gamma} 
\right) 
\nonumber\\
\hat{\cal V}_{\Gamma} &=& \frac{1}{N} \sum_\gamma   
\hat T^{\Gamma}_{0,\gamma} \, \hat T^{\Gamma}_{0,\gamma} ~,
\ea
The five interaction terms in Eq.~(10) correspond 
to  linear combinations of 
 the ones obtained in lowest order in \cite{seis,trece}.  
However, it is necessary  to include interactions  which are related 
to the vibrational angular momenta associated with the 
degenerate modes $E$ and $F_2$. These kind of terms is absent in the 
former versions of the model \cite{seis,trece}.
We now proceed to show how they can be obtained in the present model.  
In configuration space the vibrational angular momentum operator 
for the $E$ mode is given by \cite{catorce}
\ba
\hat l^{A_2} &=& -i \left( q^E_1 \, \frac{\partial}{\partial q^E_2}
- q^E_2 \, \frac{\partial}{\partial q^E_1} \right) ~,
\ea
where $q^E_1$ and $q^E_2$ are the  normal coordinates
associated to the  $E$ mode.   
This relation can be transformed to the algebraic space by means of 
the harmonic oscillator operators
\ba
b^{\Gamma \, \dagger}_{\gamma} \;=\; \frac{1}{\sqrt{2}} \left( 
q^{\Gamma}_{\gamma} - \frac{\partial}{\partial q^{\Gamma}_{\gamma}}
\right) ~, \hspace{1cm}
b^{\Gamma}_{\gamma} \;=\; \frac{1}{\sqrt{2}} \left( 
q^{\Gamma}_{\gamma} + \frac{\partial}{\partial q^{\Gamma}_{\gamma}}
\right) ~,
\ea
to obtain
\ba
\hat l^{A_2} &=& -i \left( b^{E \, \dagger}_1 b^{E}_2 - 
b^{E \, \dagger}_2 b^{E}_1 \right) ~.
\label{vibang}
\ea
Here $b^{E}_{\gamma} = \sum_{i} \alpha^E_{\gamma,i} \, b_i$, with a
similar form for $b^{\Gamma \, \dagger}_{\gamma}$, while the
$\alpha^E_{\gamma \, i}$ can be read from (4).   
In order to find the algebraic expression for $\hat l^{A_2}$  
 we first  introduce a scale transformation in (11) 
\ba
\bar b^{\dagger}_i \;\equiv\; \hat J_{-,i}/\sqrt{N_i} ~,
\hspace{1cm} 
\bar b_i \;\equiv\; \hat J_{+,i}/\sqrt{N_i} ~. \label{subst}
\ea
The relevant commutator  can be expressed as
\ba
[\bar b_i, \bar b_i^\dagger] \;=\; 
\frac{1}{N_i} [ \hat J_{+,i},\hat J_{-,i}] 
\;=\; \frac{1}{N_i} 2\hat J_{0,i} \;=\; 1 - \frac{2 \hat v_i}{N_i} ~,
\ea
where 
\ba
\hat v_i=\frac{\hat N_i}{2}-\hat J_{0,i} ~.
\ea
The other
two commutators in (11) are not modified by (11).  
In the harmonic limit, which is defined by $N_i \rightarrow \infty$,
 Eq. (12)  
reduces to the standard boson commutator  $[\bar b_i, \bar
b^\dagger_i]=1$.  
This limit corresponds to a contraction of $SU(2)$ to the Weyl algebra
and can be used to obtain a geometric interpretation of the algebraic
operators in terms of those in configuration space. 
In the opposite sense, Eq.~(5) provides a procedure to 
construct the  anharmonic representation of harmonic operators 
through the correspondence 
$b^{\dagger}_i \rightarrow \bar b^\dagger_i = 
 \hat J_{-,i}/\sqrt{N_i}$ and 
$b_i \rightarrow \bar b _i = \hat J_{+,i}/\sqrt{N_i}$.
Applying this method to the vibrational angular momentum 
(4) we find 
\ba 
\hat l^{A_2} &=& -\frac{i}{N} \left( \hat J^E_{-,1} \hat J^E_{+,2} -
\hat J^E_{-,2} \hat J^E_{+,1} \right) ~.
\ea
For the vibrational angular momentum $\hat l^{F_1}_{\gamma}$ 
associated with the $F_2$ mode we find a similar 
expression. 

\begin{table}
\centering
\caption[]{
Vibrational excitations of Be$_4$ using the algebraic Hamiltonian 
with parameters given in the text. The ab initio 
($N \rightarrow \infty$) spectrum is generated with the parameters 
from \cite{quince}. The energies are given in cm$^{-1}$.
}
\vspace{10pt} 
\footnotesize
\begin{tabular}{cclcc|cclcc}
\hline
& & & & & & & & & \\
$V$ & $(\nu_1,\nu_2^m,\nu_3^l)$ & $\Gamma$ 
& Ab initio & Present &
$V$ & $(\nu_1,\nu_2^m,\nu_3^l)$ & $\Gamma$ 
& Ab initio & Present \\
& & & $N \rightarrow \infty$ & $N=44$ & 
& & & $N \rightarrow \infty$ & $N=44$ \\
& & & & & & & & & \\
\hline
& & & & \\
1 & $(1,0^0,0^0)$     & $A_1$ &  638.6 &  637.0 & 
3 & $(1,0^0,2^0)$     & $A_1$ & 2106.8 & 2105.6 \\
  & $(0,1^1,0^0)$     & $E$   &  453.6 &  455.0 & 
  & $(1,0^0,2^2)$     & $E$   & 2000.1 & 1999.8 \\
  & $(0,0^0,1^1)$     & $F_2$ &  681.9 &  678.2 & 
  &                   & $F_2$ & 2056.8 & 2052.8 \\
2 & $(2,0^0,0^0)$     & $A_1$ & 1271.0 & 1269.2 & 
  & $(0,3^1,0^0)$     & $E$   & 1341.3 & 1343.7 \\
  & $(1,1^1,0^0)$     & $E$   & 1087.1 & 1087.0 & 
  & $(0,3^3,0^0)$     & $A_1$ & 1355.5 & 1352.5 \\
  & $(1,0^0,1^1)$     & $F_2$ & 1312.6 & 1308.3 & 
  &                   & $A_2$ & 1355.5 & 1354.4 \\
  & $(0,2^0,0^0)$     & $A_1$ &  898.3 &  901.4 & 
  & $(0,2^{0,2},1^1)$ & $F_2$ & 1565.5 & 1565.7 \\
  & $(0,2^2,0^0)$     & $E$   &  905.4 &  906.1 & 
  &                   & $F_2$ & 1584.4 & 1583.1 \\
  & $(0,1^1,1^1)$     & $F_1$ & 1126.7 & 1125.1 & 
  & $(0,2^2,1^1)$     & $F_1$ & 1578.5 & 1578.0 \\
  &                   & $F_2$ & 1135.5 & 1134.1 & 
  & $(0,1^1,2^{0,2})$ & $E$   & 1821.4 & 1821.6 \\
  & $(0,0^0,2^0)$     & $A_1$ & 1484.0 & 1483.0 & 
  &                   & $E$   & 1929.5 & 1929.0 \\
  & $(0,0^0,2^2)$     & $E$   & 1377.3 & 1373.9 & 
  & $(0,1^1,2^2)$     & $A_2$ & 1813.3 & 1813.1 \\
  &                   & $F_2$ & 1434.1 & 1429.6 & 
  &                   & $A_1$ & 1830.8 & 1831.7 \\
3 & $(3,0^0,0^0)$     & $A_1$ & 1897.0 & 1896.7 & 
  &                   & $F_2$ & 1874.4 & 1873.2 \\
  & $(2,1^1,0^0)$     & $E$   & 1714.3 & 1714.3 & 
  &                   & $F_1$ & 1883.2 & 1883.0 \\
  & $(2,0^0,1^1)$     & $F_2$ & 1937.0 & 1933.7 & 
  & $(0,0^0,3^{1,3})$ & $F_2$ & 2136.5 & 2134.2 \\
  & $(1,2^0,0^0)$     & $A_1$ & 1526.6 & 1529.2 & 
  &                   & $F_2$ & 2327.3 & 2326.9 \\
  & $(1,2^2,0^0)$     & $E$   & 1533.7 & 1532.8 & 
  & $(0,0^0,3^3)$     & $F_1$ & 2199.8 & 2197.1 \\
  & $(1,1^1,1^1)$     & $F_1$ & 1752.2 & 1749.7 & 
  &                   & $A_1$ & 2256.5 & 2254.4 \\
  &                   & $F_2$ & 1761.0 & 1759.8 & 
  &                   &       &        &        \\
& & & & & & & & & \\
\hline
\end{tabular}
\normalsize
\end{table}

We can now use  
our model to fit the spectroscopic data of several polyatomic
molecules.  In the case of Be$_4$ the energy spectrum  
was analyzed by {\it ab initio} methods in \cite{quince}, where force-field
constants corresponding to an expansion of the potential up to fourth
order in the normal coordinates and momenta were evaluated. We have
generated the {\it ab initio} spectrum up to three phonons using the
analysis in \cite{catorce}. For the algebraic Hamiltonian we take 
\cite{siete}
\ba
\hat H &=& \omega_1 \, \hat{\cal H}_{A_1} 
+ \omega_2 \, \hat{\cal H}_{E} + \omega_3 \, \hat{\cal H}_{F_2} 
+ X_{33} \left( \hat{\cal H}_{F_2} \right)^2
+ X_{12} \left( \hat{\cal H}_{A_1} \hat{\cal H}_{E  } \right)
\nonumber\\
&& + X_{13} \left( \hat{\cal H}_{A_1} \hat{\cal H}_{F_2} \right)
+ g_{33} \, \sum_{\gamma} \hat l^{F_1}_{\gamma} \, \hat l^{F_1}_{\gamma} 
+ t_{33} \, \hat{\cal O}_{33} + t_{23} \, \hat{\cal O}_{23} ~.
\ea
The terms $\hat{\cal O}_{33}$ and $\hat{\cal O}_{23}$ represent 
the algebraic form of the corresponding interactions 
in \cite{catorce} which are responsible for the splitting of the 
vibrational levels in the $(\nu_1,\nu_2^m,\nu_3^l)=(0,0^0,2^2)$ and 
the $(0,1^1,1^1)$ overtones.  

Note that the Be$_4$ Hamiltonian (15) preserves the total
number of quanta $V$.  This is a good approximation for this case
according to the analysis of \cite{catorce,quince}, but it is
known that Fermi  
resonances can occur for certain molecules when the fundamental mode
frequencies are such that $(V,V^\prime)$ states with $V \neq
V^\prime$ are close in energy.  These interactions can be introduced
in the Hamiltonian  by means of a polyad analysis \cite{dieciseis}. 
For ${\cal D}_{3h}$ molecules we
can follow an analogous procedure, namely, we can construct the
${\cal D}_{3h}$ symmetry-adapted operators and states corresponding
to (4) and \cite{siete} and carry out the building up procedure to construct
the Hamiltonian and higher phonon states, using in this case the
appropriate projection operators and Clebsch-Gordan
coefficients \cite{seis,once}.  

\begin{table}
\centering
\caption[]{Least-square energy fit for the vibrational 
excitations of H$^+_3$, Be$_3$ and Na$^+_3$. The energy differences 
$\Delta E = E_{th} - E_{exp}$ are given in cm$^{-1}$.
}
\vspace{10pt}
\footnotesize
\begin{tabular}{cccrrr}
\hline
& & & & & \\
& & & H$^+_3$ & Be$_3$ & Na$^+_3$ \\
$V$ & $(\nu_1,\nu_2^l)$ & $\Gamma$ & $\Delta E$ & $\Delta E$ & $\Delta E$ \\ 
& & & & & \\
\hline
& & & & & \\
1 & $(0,1^1)$ & $E$   & -1.55 &  0.51 &  0.93 \\ 
  & $(1,0^0)$ & $A_1$ &  0.42 &  0.02 &  1.95 \\ 
& & & & &  \\
2 & $(0,2^0)$ & $A_1$ &  7.48 & -0.74 &  0.37 \\ 
  & $(0,2^2)$ & $E$   & -5.69 &  0.17 &  0.84 \\
  & $(1,1^1)$ & $E$   & -0.61 &  0.82 &  1.68 \\
  & $(2,0^0)$ & $A_1$ & -0.11 & -0.04 &  1.26 \\
& & & & &  \\ 
3 & $(0,3^1)$ & $E$   & -4.46 & -2.05 & -1.19 \\ 
  & $(0,3^3)$ & $A_1$ &  3.18 & -1.23 & -0.34 \\
  & $(0,3^3)$ & $A_2$ &  2.44 &  0.61 & -0.33 \\
  & $(1,2^0)$ & $A_1$ &  0.66 &  1.90 & -0.01 \\
  & $(1,2^2)$ & $E$   & -5.00 & -1.36 &  0.34 \\
  & $(2,1^1)$ & $E$   &  4.07 &  0.79 & -0.19 \\
  & $(3,0^0)$ & $A_1$ & -1.23 & -1.66 & -2.06 \\
& & & & & \\
\hline 
& & & & & \\ 
& & r.m.s. &  5.84 &  1.35 &  1.33 \\
& & & & & \\
\multicolumn{3}{r} {Parameters} & 8 & 4 & 4 \\
& & & & & \\
\hline
\end{tabular}
\normalsize
\end{table}

\vspace{8mm}
\noindent{\bf  EXAMPLES}
\vspace{6mm}

We now present the results of our least-square
fits to the energy spectra of Be$_4$, Be$_3$, Na$^+_3$ and H$^+_3$. 
  In Table~1 we show the fit to Be$_4$ using the Hamiltonian (15).  
The fit includes all levels up to $V=4$ quanta and gives a  
r.m.s. deviation of 2.6 cm$^{-1}$, which can be considered
of spectroscopic quality. In Table~1 we only show the results 
for the levels with $V \leq 3$.  We point out that in
 \cite{catorce,quince}  several
higher order interactions are present which we have  neglected.
Since our model  can be put into a one to one correspondence with the
configuration space calculations, it is in fact possible to improve
the accuracy of the fit considerably, but we have used a simpler
Hamiltonian than the one of \cite{catorce,quince}.  When no
{\it ab initio} 
calculations are available (or feasible) the present approach can be
used empirically, achieving increasingly good fits by the inclusion
of higher order interactions \cite{siete}.  In Table~2 we
present  fits to the spectra of Be$_3$, Na$^+_3$ and H$^+_3$ up
to three phonons.  While remarkably accurate descriptions of the
first two molecules can be achieved using a four-parameter
Hamiltonian, we were forced to include four additional higher order
terms in the H$^+_3$ Hamiltonian in order to properly describe this
molecule.  This is in accordance with the work of Carter and
Meyer \cite{diecisiete}, who were forced to include twice as many
terms in the potential energy surface for H$^+_3$ than for the
Na$^+_3$ molecule.  The H$^+_3$ ion is a very ``soft'' molecule
which, due to the light mass of its atomic constituents carries out
large amplitude oscillations from its equilibrium
positions \cite{diecisiete}.  This may also be the case for the
potentials associated to metallic clusters \cite{dos}.  

\vspace{8mm}
\noindent{\bf SUMMARY}
\vspace{6mm}

In this paper we have studied the vibrational excitations 
of several molecules and clusters in a symmetry-adapted algebraic model.  
 These studies suggest that the symmetry-adapted algebraic model 
provides a numerically efficient tool to study molecular 
vibrations with high precision. The main difference with other 
methods is the use of symmetry-adapted tensors in the construction 
of the Hamiltonian. In this approach, the anharmonicity can be
introduced from the outset,  interactions can be constructed 
in a systematic way, each term has a direct physical interpretation, 
and spurious modes can be eliminated exactly \cite{dieciseis}.   
It is important to further explore the 
scope and applicability of the present approach, and in particular 
 study  the vibrational spectra in metallic clusters, including the
appearance of isomeric shapes with different point symmetries.

\vspace{8mm}
\noindent{\bf ACKNOWLEDGEMENTS}
\vspace{6mm}

This work was supported in part by the 
European Community under contract nr.~CI1$^{\ast}$-CT94-0072, 
DGAPA-UNAM under project IN101997 and Spanish DGCYT under project
PB92-0663.

\vfill
\eject



\end{document}